%% file: note.tex
\documentclass[a4paper,12pt]{article}
\input{definitions}
\usepackage[noblocks]{authblk}

\RequirePackage[normalem]{ulem} 
\RequirePackage{color}\definecolor{RED}{rgb}{1,0,0}\definecolor{BLUE}{rgb}{0,0,1} 

\begin{document}

\makeatletter
\begin{titlepage}

\includegraphics[width=3cm]{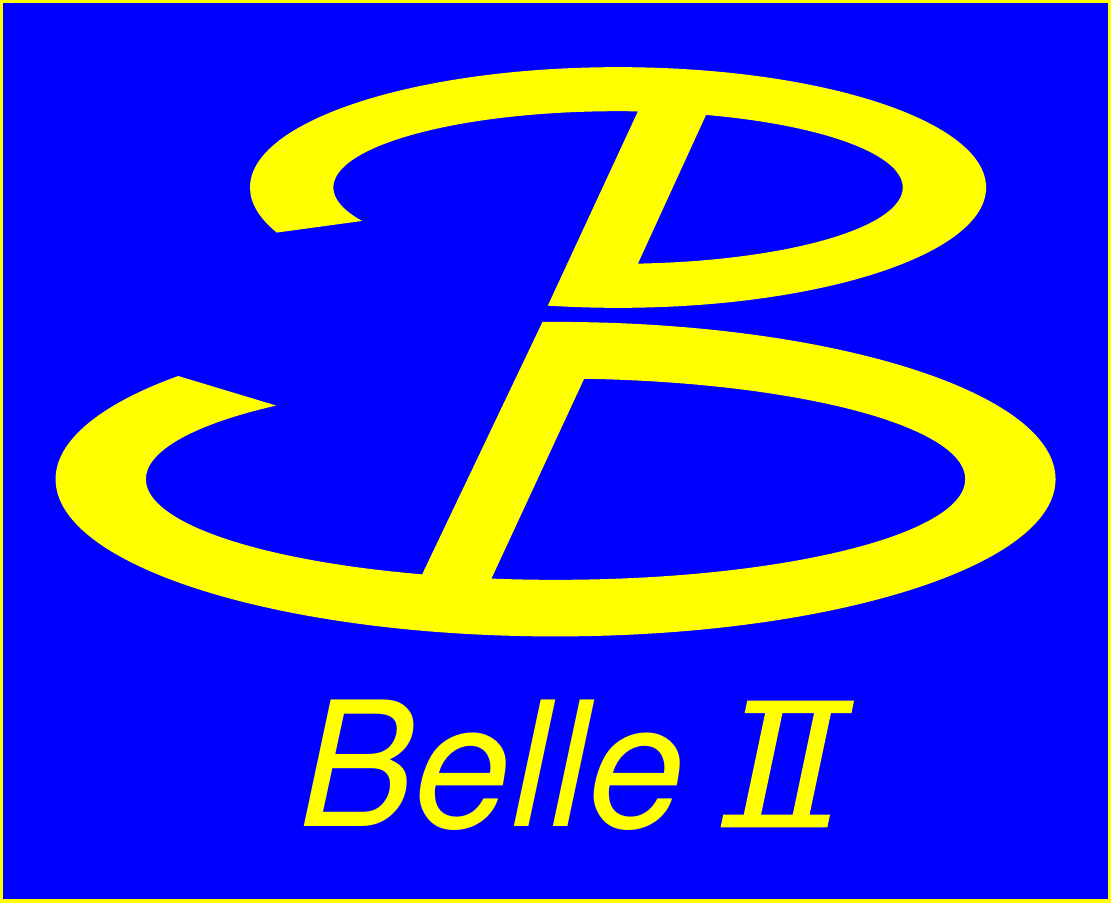}\vspace*{-3cm}

\begin{flushright}
\vspace*{24pt}
BELLE2-CONF-2022-016 \\ 
\today
\end{flushright}
\vspace*{24pt}

\centering
{\Large Measurement of Branching Fraction and Longitudinal Polarization in $B^0 \to \rho^+ \rho^-$ Decays at Belle II}\\[24pt] 

\input{authors.tex}

\@author\\[12pt]

\textit{The Belle II Collaboration}\\[24pt]

\begin{abstract}
We present a measurement of the branching fraction and longitudinal polarization of $B^0 \to \rho^+ \rho^-$ decays.
SuperKEKB electron-positron collision data corresponding to 189~fb$^{-1}$ of integrated luminosity and containing $198 \times 10^6 B\bar{B}$ pairs collected with the Belle II detector are used.
We obtain 
\begin{eqnarray*}
  \mathcal{B}(B^0\to\rho^+\rho^-) &=& [2.67\pm0.28\,(\mathrm{stat})\,\pm0.28\,(\mathrm{syst})] \times 10^{-5},\\
  f_L &=& 0.956\pm0.035\,(\mathrm{stat})\,\pm 0.033\,(\mathrm{syst}),
\end{eqnarray*}
These results are consistent with previous measurements and can be used to constrain penguin pollution and to extract the quark-mixing angle $\phi_2$. 
\end{abstract}

\end{titlepage}
\makeatother


\input{main.tex}

\section*{Acknowledgements}
We thank the SuperKEKB group for the excellent operation of the
accelerator; the KEK cryogenics group for the efficient
operation of the solenoid; the KEK computer group for
on-site computing support.

\FloatBarrier

\newpage
\bibliographystyle{./belle2-note.bst}
\bibliography{belle2}

\end{document}

%% file: definitions.tex
\usepackage{graphicx} 
\usepackage{epstopdf} 
\usepackage{dcolumn} 
\usepackage{xcolor} 
\usepackage{amssymb} 
\usepackage{hyperref} 
\usepackage[utf8]{inputenc} 
\usepackage[english]{babel} 
\usepackage[displaymath, mathlines]{lineno} 
\usepackage{blindtext} 
\usepackage{amsmath,bm}
\usepackage{placeins}

\graphicspath{{figures/}} 

%% file: authors.tex
  \author{F.~Abudin{\'e}n,  
  I.~Adachi,  
  K.~Adamczyk,  
  L.~Aggarwal,  
  P.~Ahlburg,  
  H.~Ahmed,  
  J.~K.~Ahn,  
  H.~Aihara,  
  N.~Akopov,  
  A.~Aloisio
% ...
}

%% file: main.tex

\section{Introduction}
Violations of charge-parity ({\it CP}) symmetry in the Standard Model~(SM) are described by a single irreducible complex phase in the Cabibbo-Kobayashi-Maskawa~(CKM) quark-mixing matrix~\cite{Cabibbo:1963yz,Kobayashi:1973fv}.
The Belle and BaBar experiments have observed {\it CP} violation in the $B$ meson system~\cite{Belle:2001zzw,Belle:2002bwy,BaBar:2001pki,BaBar:2002epc} and measured three angles and three sides of the unitarity triangle~(UT).
These UT observables consistently point to a single apex with a precision of ${\mathcal{O}}$(10)\%.
This verifies the CKM picture of {\it CP} violation in the SM quantitatively and provides stringent constraints on non-SM physics contributions.
However, there is still room for a non-SM physics amplitude that is ${\mathcal{O}}$(10)\% of the SM amplitude~\cite{Charles:2013aka,Tanimoto:2014eva}. 
Thus, further improvement of the precision of the UT is crucial to search for non-SM physics. 

Among the six observables of the UT, the angle $\phi_2$, also known as $\alpha$, which is expressed in terms of CKM matrix elements as ${\rm{arg}} (-V_{td}V_{tb}^*/V_{ud}V_{ub}^*)$, is the least precisely measured, $\phi_2 = (85.2^{+4.8}_{-4.3})^{\circ}$~\cite{ParticleDataGroup:2022zzz}.
The angle $\phi_2$ can be measured by studying the decay-time-dependent asymmetry between $B^0$ and $\bar{B}^0$ yields in $b \to u \bar{u} d$ transitions such as those governing the decays $B \to \pi \pi$, $B \to \rho \pi$, $B \to \rho \rho$ and $B \to a_1 \pi$.
These decays proceed via $b \to u$ tree amplitudes and $b \to d$ penguin amplitudes. 
The former is the leading contribution, which has a weak phase of $\phi_2$ including the phase from $B^0$-$\bar{B}^0$ mixing, while the latter contribution is subdominant and carries a weak phase different from $\phi_2$, which can bias the observed phase to $\phi_2^{{\rm eff}} \equiv \phi_2 + \Delta\phi_2$.
This is called penguin pollution.

To estimate the penguin pollution and the shift $\Delta \phi_2$, an isospin analysis using as inputs branching fractions and direct {\it CP} asymmetries in these decays is needed~\cite{Gronau:1990ka}.
From the current isospin analyses, it is known that $B^0 \to \rho^+ \rho^-$ has a small contribution from penguin amplitudes, and is the golden mode for the $\phi_2$ measurement.
To further improve the understanding of penguin pollution, precise measurements of branching fractions of $B \to \rho \rho$ modes are needed.

Since $B^0 \to \rho^+ \rho^-$ is a pseudo-scalar to vector vector decay, there are three helicity states of the $\rho$ meson pair, longitudinal polarization $H_0$ (pure {\it CP}-even), and transverse polarizations $H_+$ and $H_-$ (a mixture of {\it CP}-even and {\it CP}-odd)~\cite{CC}.
Only $H_0$ is usable for the time dependent analysis to extract the {\it CP} violation parameters.
In the factorization limit, the fraction of longitudinal polarization, defined as $f_L = |H_0|^2/(|H_0|^2 + |H_+|^2 + |H_-|^2)$, is close to unity~\cite{Kagan:2004uw}.
The $f_L$ parameter can be measured with an angular analysis of $\rho^+$ mesons described as \begin{eqnarray}
\frac{1}{\Gamma} \frac{d^2\Gamma}{d\cos\theta_{\rho^+} d\cos\theta_{\rho^-}} 
= \frac{9}{4}\Big[\frac{1}{4}(1-f_L)\sin^2\theta_{\rho^+}\sin^2\theta_{\rho^-} + f_L \cos^2\theta_{\rho^+} \cos^2\theta_{\rho^-}\Big],
\end{eqnarray}
where $\theta_{\rho^+}$ is the angle between the momentum of the daughter of the $\rho^+$ meson (the $\pi^0$ is taken in this paper) and the $B$ flight direction in the rest frame of the $\rho^+$ meson.
Knowledge of the $f_L$ value determines the sensitivity to the {\it CP} violating parameters, and thus a precise determination of $f_L$ is required.

Belle and BaBar have measured the branching fraction and longitudinal polarization in $B^0 \to \rho^+\rho^-$ as summarized in Table~\ref{tab:prev}~\cite{ParticleDataGroup:2022zzz, Belle:2015xfb, BaBar:2007cku} and confirmed the dominance of longitudinal polarization.

\begin{table}[hbt]
  \caption{Recent precise measurements and world averages for $B \to \rho^+ \rho^-$ decays.}
  \label{tab:prev}
  \centering
  \begin{tabular}{lll}
  \hline \hline
  Exp. & ${\mathcal{B}}$ [$10^{-5}$] & $f_{L}$ \\
  \hline
  Belle     & $2.83 \pm 0.15 \pm 0.15$          & $0.988 \pm 0.012 \pm 0.023$\\
  BaBar     & $2.55 \pm 0.21 {}^{+0.36}_{-0.39}$ & $0.992 \pm 0.024 {}^{+0.026}_{-0.013}$ \\
  \hline
  PDG2022   & $2.77 \pm 0.19$                  & $0.990^{+0.021}_{-0.019}$ \\
  \hline \hline
  \end{tabular}
\end{table}

In this paper, we present a measurement of the branching fraction and longitudinal polarization in $B \to \rho^+ \rho^-$ based on a data sample corresponding to 189~fb$^{-1}$ of integrated luminosity and containing $(198 \pm 3) \times 10^6 B\bar{B}$ pairs collected on the $\Upsilon(4S)$ resonance with Belle~II detector at the SuperKEKB energy-asymmetric $e^+e^-$ collider.

\section{Belle II detector}
Belle II is a nearly $4\pi$ particle-physics spectrometer~\cite{Kou:2018nap, Abe:2010sj}, designed to reconstruct the products of electron-positron collisions produced by the SuperKEKB asymmetric-energy collider~\cite{Akai:2018mbz}, located at KEK in Tsukuba, Japan.
Belle II comprises several subdetectors arranged around the interaction point in a cylindrical geometry.
The innermost subdetector is the vertex detector, which uses position-sensitive silicon layers to sample the trajectories of charged particles (tracks) in the vicinity of the interaction region to extrapolate the decay positions of their long-lived parent particles.
The vertex detector includes two inner layers of silicon pixel sensors and four outer layers of silicon microstrip sensors.
The second pixel layer is currently incomplete and covers only one-sixth of the azimuthal angle. 
Charged-particle momenta and charges are measured by a large-radius, helium-ethane, small-cell central drift chamber (CDC), which also offers charged-particle-identification information through measurement of particles’ energy loss by specific ionization.
A time-of-propagation Cherenkov detector surrounding the chamber provides charged-particle identification (PID) in the central detector volume, supplemented by proximity-focusing, aerogel ring-imaging Cherenkov detectors in the forward regions.
A CsI(Tl)-crystal electromagnetic calorimeter (ECL) allows for energy measurements of electrons and photons.
A solenoid surrounding the calorimeter generates a uniform axial 1.5~T magnetic field filling its inner volume.
Layers of plastic scintillators and resistive-plate chambers, interspersed between the magnetic flux-return iron plates, allow for identification of $K_L^0$ and muons.
The subdetectors most relevant for this work are the silicon vertex detector, drift chamber, the PID detectors, and the electromagnetic calorimeter.

\subsection{Data and simulation}
We use all 2019-2021 data corresponding to an integrated luminosity of 189~fb$^{-1}$ collected with the Belle II detector.
Data are collected at the center-of-mass (CM) energy of the $\Upsilon (4S)$ resonance ($\sqrt{s} =10.58~\mathrm{GeV}$).
The energies of the electron and positron beams are 7~GeV and 4~GeV, respectively, resulting in a boost factor of $\beta \gamma = 0.28$ of the CM frame relative to the laboratory frame.
We also use all off-resonance data collected at an energy about 60~MeV lower, which corresponds to an integrated luminosity of 18.0~$\mathrm{fb^{-1}}$.
All events are required to satisfy loose hadronic event selection criteria, based on total energy and neutral-particle multiplicity in the events, targeted at reducing sample sizes to a manageable level with minimal impact on signal efficiency.
All data are processed using the Belle II analysis software~\cite{Kuhr:2018lps}.

We use simulated data based on GEANT4~\cite{Agostinelli:2002hh} Monte Carlo (MC) to optimize the event selection, compare the distributions observed in experimental data with expectations, and model the distributions in fits.
A sample of $2\times10^5$ signal-only $B^0\to\rho^+\rho^-$ events are used for each polarization.
Generic background MC samples consist of charged and neutral $B$ meson pairs ($B^0\bar{B}^0$ and $B^+B^-$), and continuum processes ($e^+e^- \to q\bar{q}$ with $q$ = $u$, $d$, $s$, $c$ quarks) in realistic proportions, and correspond to a 1~$\mathrm{ab^{-1}}$ sample.
In addition, we generate $5\times10^5$ events of simulated data for each peaking background whose final state is the same as or similar to $B^0\to\rho^+\rho^-$ decay.
To validate our experimental procedure, the $B^- \to D^0 (\to K^- \pi^+ \pi^0) \rho^- (\to \pi^- \pi^0)$ decay is used as a control mode as it contains a $\pi^+$, $\pi^-$, and two $\pi^0$'s in the final state.
Finally, simplified simulated experiments constructed by randomly sampling the likelihood of the sample-composition fit allow us to study the estimator properties and assess fit-related systematic uncertainties.

\section{Reconstruction and selection}
We reconstruct the two-body decay $B^0 \to \rho^+ (\to \pi^+ \pi^0 (\to \gamma\gamma)) \rho^- (\to \pi^- \pi^0 (\to \gamma\gamma))$ while using appropriately selected particle candidates at each reconstruction step. 
To reliably and efficiently search for the best selection criteria, we use a genetic optimization algorithm, differential evolution~\cite{DE}, which iteratively changes the selection criteria for multiple selection parameters and maximizes the figure of merit $S/\sqrt{S+B}$, where $S$ and $B$ are the simulated signal and background yields, respectively.

\begin{figure}[t]
  \centering
  \includegraphics[width=0.7\textwidth]{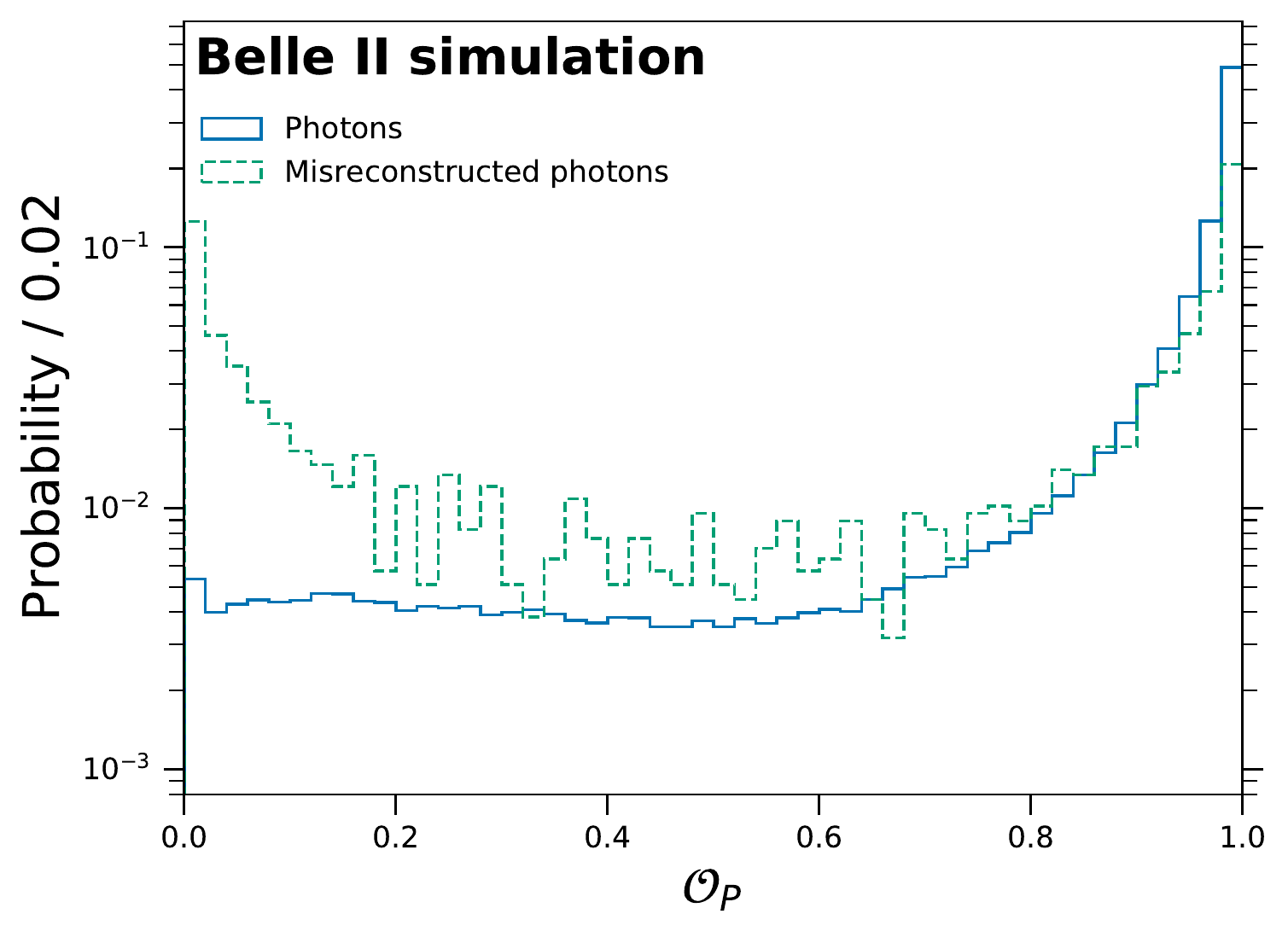}
  \caption{Distribution of the FBDT classifier to distinguish correctly reconstructed photons from hadronic clusters and split-off showers from charged-particle tracks in simulation.}
  \label{fig:pmva}
\end{figure}

Track candidates must be in the full polar-angle acceptance of the CDC ($17^\circ < \theta<150^\circ$) and satisfy requirements on the distance of closest approach to the interaction point to reject misreconstructed tracks and charged particles from beam background, $|dr|<0.5~\mathrm{cm}$ in the radial direction and $|dz|<3.0~\mathrm{cm}$ in the longitudinal direction. We select charged pions from the track candidates using loose criteria based on PID information.
Photons are identified as groups of adjacent ECL channels having signal above threshold (clusters) and not matched to tracks.
For photon candidates, the minimum energy is set to 50~MeV, 60~MeV, and 100~MeV in the ECL barrel, forward endcap, and backward endcap with respect to the CM boost direction, respectively, in order to suppress combinatorial $\pi^0$ misreconstruction due to contamination of low-energy photons.
In addition, selections on ECL-cluster timing and the number of ECL crystals detecting a signal are applied.
To distinguish between correctly reconstructed photons and photons misreconstructed from hadronic clusters and showers splitting from hadronic clusters (split-off showers), we train a fast boosted decision-tree (FBDT)~\cite{Belle-II:FBDT} with 14 variables associated with cluster shapes and uncertainties of the cluster information, as shown in Figure~\ref{fig:pmva}.
We choose the threshold on the FBDT output, $\mathcal{O}_P>0.1$, that maximizes the figure of merit.
We reconstruct neutral-pion candidates by combining pairs of photons whose invariant mass is near the $\pi^0$ mass, $118~\mathrm{MeV}/c^2<m_{\gamma\gamma}<151~\mathrm{MeV}/c^2$.
In addition, $\pi^0$ candidates must have momenta greater than $240~\mathrm{MeV}/c$, with the angle between the two photon directions less than 0.1 radians in $\theta$ and less than 2.2 radians in $\phi$.
The selected $\pi^+$ and $\pi^0$ candidates are combined to form dipion-state ($\rho^+$) candidates by requiring $0.6~\mathrm{GeV}/c^2<m_{\pi^+\pi^0}<1.1~\mathrm{GeV}/c^2$.
The resulting $B$ signal yield is determined using two kinematic variables: the beam-energy-constrained mass $M_{\rm{bc}}\equiv \sqrt{(E_{\rm{beam}}^{*})^2-(p_{B}^{*})^2}$ and the energy difference $\Delta E \equiv E_{B}^{*} - E_{\rm{beam}}^{*}$, where $E_{\rm{beam}}^{*}$ is the beam energy and $E_{B}^{*}$ ($p_{B}^{*}$) is the energy (momentum) of the $B$ meson, evaluated in the CM frame.
Candidate $B$ mesons are required to have $M_{\rm{bc}}>5.274~\mathrm{GeV}/c^2$ and $|\Delta E|<0.15~\mathrm{GeV}$.

\begin{figure}[t]
  \centering
  \includegraphics[width=0.7\textwidth]{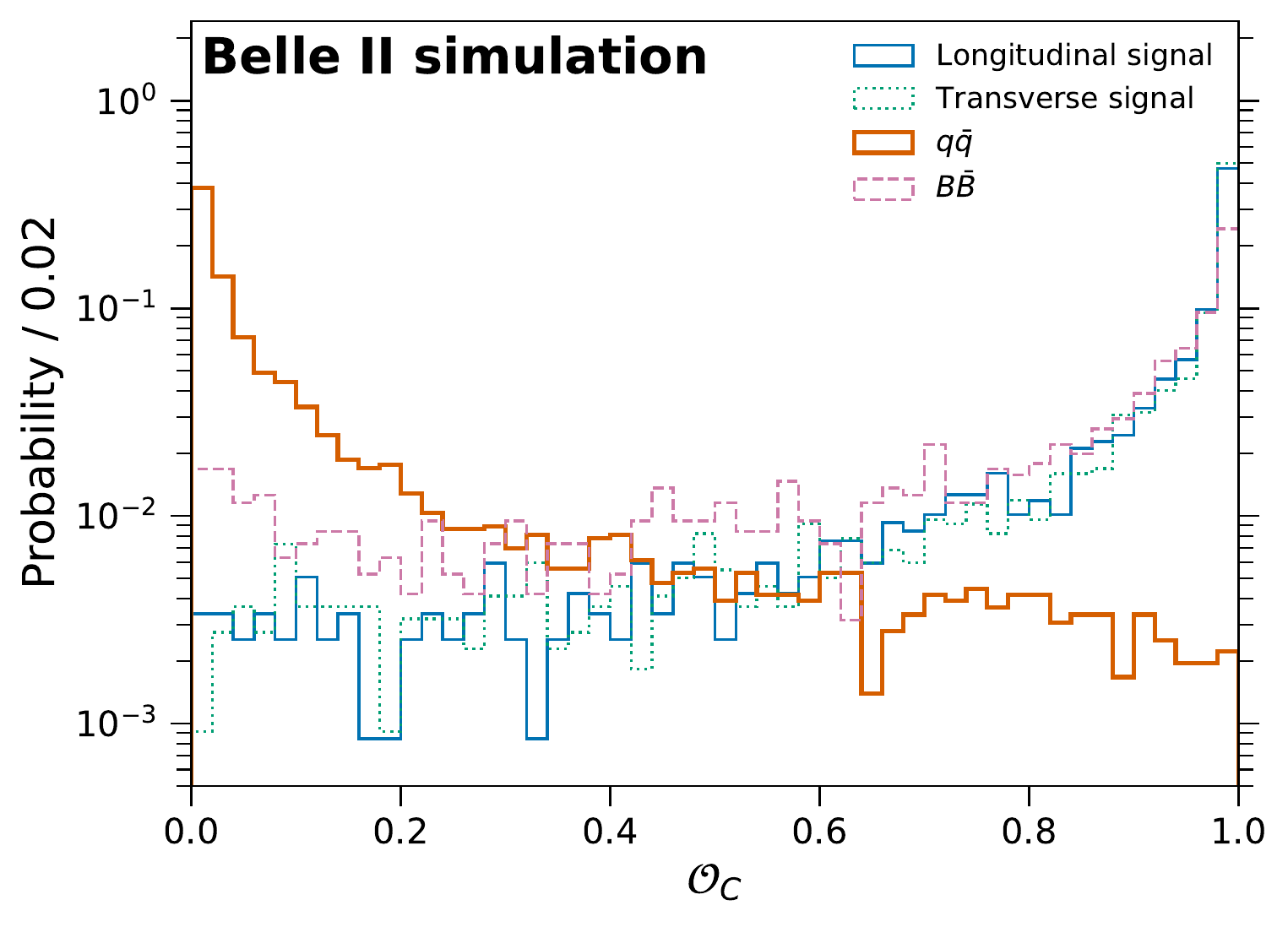}
  \caption[Distribution of the FBDT classifier for continuum suppression in a MC simulation]{Distribution of the FBDT classifier for continuum suppression in a MC simulation.}
  \label{fig:csmva}
\end{figure}

The large continuum background is suppressed further to observe charmless $B$ decays.
To discriminate against continuum processes, we use an FBDT classifier that combines 28 variables associated with event topology, which are known to provide statistical discrimination between a $B$-meson signal and continuum background.
We train the classifier to identify statistically significant signal and background features using unbiased simulated samples.
We validate the input and output distributions of the classifier by comparing data with simulation using the control mode.
Figure~\ref{fig:csmva} shows the distribution of the FBDT-classifier output for $B^0\to D^{*-} (\to \bar{D}^0 (\to K^+\pi^-) \pi^-) \pi^+$ candidates.
The selection criterion on the continuum-suppression output is set to $\mathcal{O}_C>0.98$, which removes 97\% of continuum and retains 54\% of signal.

More than one candidate per event often populates the resulting sample, with an average multiplicity of 1.3.
We restrict to one candidate per event by selecting the candidate with the minimum sum of reduced-$\chi^2$ for the $\pi^0$ mass-constrained diphoton fit and the $B$-vertex fit.
After the optimized selection, 94.9\% and 98.3\% of the remaining longitudinally- and transversely-polarized signal events are correctly reconstructed.

\section{Signal yield extraction}
Signal yields are determined with a six-dimensional ($\Delta E$, $m_{\pi^\pm\pi^0}$, $\cos \theta_{\rho^\pm}$, $T_C$) extended unbinned maximum likelihood fit.
Here $T_C$ is a variable converted from continuum suppression FBDT output $\mathcal{O}_C$ using an ordered list of $\mathcal{O}_C$ values from the signal simulation sample such that $T_C$ is the fraction of signal events present below a given value of $\mathcal{O}_C$ in the list~\cite{Hocker:2007ht}.
Therefore, $T_C$ is distributed uniformly for the signal, while it peaks at zero for the continuum background.

The fit region is defined as follows: $\Delta E \in (-0.15,~0.15)~\mathrm{GeV}$, $m_{\pi^\pm\pi^0} \in (0.6,~1.1)~\mathrm{GeV}/c^2$, $\cos \theta_{\rho^\pm} \in (-1,~0.9)$, and $T_C \in (0,~1)$.
Sample components used in the fit are longitudinally-polarized signal, transversely-polarized signal, their self cross-feed components (i.e., incorrectly reconstructed candidates in signal events), $B^0\to \rho^+ \pi^-\pi^0$, $B^0\to\pi^+\pi^0\pi^-\pi^0$, $B^0\to a^0_1 \pi^0$, $B\bar{B}$ background, and continuum, whose yields are determined by the fit.
The yields of $B^0\to a^+_1\pi^-$, $B^0\to K^{*+} \rho^-$, and $B^0\to K^{*+}_0 \rho^-$ are fixed to the world-average values from~\cite{ParticleDataGroup:2022zzz}.

Fit models to construct probability density functions~(PDF) are physics-based when appropriate (e.g., dipion masses) or empirical otherwise, generally determined by parameterizing distributions from simulation using analytical functions.
Since non-negligible correlations between $\Delta E$ and $\cos \theta_{\rho^\pm}$ are seen for signal and cross-feed, the $\cos \theta_{\rho^\pm}$ distributions are parameterized as functions of $\Delta E$ while the $\Delta E$ distribution is fitted without considering the correlations.
To take into account a possible mismodeling of the $\Delta E$ shape for $B\bar{B}$ background, the relevant shape parameters, which are the coefficients of the associated second-order Chebyshev polynomial~($c_1,c_2$), are determined from the fit in data.
Table~\ref{tab:fit-models} summarizes the fit models for each sample component.
%
\begin{table}[t]
  \caption{Summary of fit models for each sample component. The acronym bG indicates a bifurcated-Gaussian, BW a Breit-Wigner formula, CX an X-th order Chebyshev polynomial function, L a Linear function, C a constant, and d double. $|_{\Delta E}$ implies that the correlation with $\Delta E$ is considered.}
  \label{tab:fit-models}
  \centering
  \begin{tabular}{lcccc}
    \hline \hline
      & $\Delta E$ & $m_{\pi^\pm\pi^0}$ & $T_C$ &$\cos \theta_{\rho^\pm}$ \\
    \hline
    Longitudinal signal~ & ~dbG~&~BW~&~C~&~Template$|_{\Delta E}$\\
    Long. self cross-feed~ & ~bG~&~BW + C1~&~L~&~Template$|_{\Delta E}$\\
    Transverse signal~ & ~dbG~&~BW~&~L~&~Template$|_{\Delta E}$\\
    Trans. self cross-feed~ & ~bG~&~BW + C1~&~L~&~Template$|_{\Delta E}$\\
    $q\bar{q}$~ & ~C2~&~BW + C1~&~exp$(ax+bx^{2})$~&~Template\\
    $B\bar{B}$~ & ~C2~&~BW + C1~&~L~&~Template\\
    $B^0\to \rho^{+} \pi^{-} \pi^{0}$~ & ~dbG~&~BW + C1~&~C1~&~Template\\
    $B^0\to \pi^{+} \pi^{-} \pi^{0} \pi^{0}$~ & ~dbG~&~C1~&~C1~&~Template\\
    $B^0\to a_{1}^{+} \pi^{-}$~ & ~bG + C1~&~C1~&~C1~&~Template\\
    $B^0\to a_{1}^0 \pi^{0}$~ & ~bG + C1~&~C2~&~C1~&~Template\\
    $B^0\to K^{*+}\rho^{-}$~ & ~dbG + C2~&~BW + C1~&~C2~&~Template\\
    $B^0\to K_{0}^{*+}\rho^{-}$~ & ~dbG + C2~&~BW + C1~&~C2~&~Template\\
    \hline \hline
  \end{tabular}
\end{table}
%
The total likelihood used for the fit to extract ${\mathcal{B}}$ and $f_L$ is 
\begin{eqnarray*}
  &\,& \mathcal{L}({\mathcal{B}}, f_L, N_{B\bar{B}}, N_{q\bar{q}}, N_{\rho^+\pi^-\pi^0}, N_{\pi^+\pi^-\pi^0\pi^0}, N_{a_1^0\pi^0},c_1,c_2 | \Delta E,~m_{\pi^\pm\pi^0},~\cos \theta_{\rho^\pm},~T_C ) \\
  &\equiv& \frac{\prod_{j} e^{-N_j}}{N_{\rm{tot}}!} \prod_{i=1}^{N_{\rm{tot}}} \sum_j N_j \mathcal{P}_j^{\Delta E} \times \mathcal{P}_j^{m_{\pi^\pm\pi^0}}\times\mathcal{P}_j^{\cos \theta_{\rho^\pm}}\times\mathcal{P}_j^{T_C},
\end{eqnarray*}
where $i$ and $j$ run over events and fit components, respectively.
Here $N_{\rm{tot}}$ is the total number of events used in the fit, $N_j$ is the yield of component $j$, and $\mathcal{P}_j$ is its PDF for the observables $\Delta E$, $m_{\pi^\pm\pi^0}$, $\cos \theta_{\rho^\pm}$, and $T_C$.
The yields $N_j$ associated with the signal decay modes are $N_{L}$, $N_{T}$, $N_{L}^{{\mathrm {SCF}}}= k_{L}^{{\mathrm {SCF}}}N_{L}$, and $N_{T}^{{\mathrm {SCF}}}= k_{T}^{{\mathrm {SCF}}}N_{T}$,
where the subscripts $L$ and $T$ denote longitudinally-polarized and transverse-polarized signal yields, respectively, the superscript SCF indicates self cross-feed, 
and $k$ is the self-cross-feed fraction with respect to the signal yield determined in simulation.
The signal yields, $N_T$ and $N_L$ are described through the branching fraction and the fraction of longitudinally-polarized decays, using the relations,
\begin{eqnarray*}
  \mathcal{B} &=& \frac{ n_L + n_T}{2f_{00}N_{B\bar{B}}\mathcal{B}_{\pi^0}^2 }, \\
  f_L &=& \frac{ n_L }{ n_L + n_T },
\end{eqnarray*}
where $n_L$ and $n_T$ are the efficiency-corrected signal yields~($n_i = N_i/\epsilon_i$, where $\epsilon$ is the efficiency) for longitudinal and transverse polarization, respectively, and where
$\mathcal{B}_{\pi^0}=(98.823\pm0.034)\%$ is the $\pi^0\to\gamma\gamma$ branching fraction~\cite{ParticleDataGroup:2022zzz}, $f_{00}=0.486\pm0.006$ is the branching fraction of $\Upsilon(4S) \to B^0 \bar{B}^0$, and $N_{B\bar{B}}=(198\pm3)\times10^6$ is the number of produced $B\bar{B}$ pairs.

\section{Systematic uncertainties} \label{sec:syst}
We evaluate the systematic uncertainties associated with particle detection efficiencies, selections, physics parameters, peaking background, modeling of signal and background PDFs, fitter bias, and MC simulation sample size.
The systematic uncertainties are listed in Table~\ref{tab:syst-error}.

We assign a systematic uncertainty of 0.3\% for each charged-particle track in the final state.
The efficiencies for longitudinal and transverse signals are determined from simulation.
The differences in efficiencies between data and simulation, and their uncertainties, are evaluated with various control samples.
The $\pi^0$ efficiency is studied using $D^0\to K^- \pi^+\pi^0$, $D^0\to K^-\pi^+$, and $B^-\to \pi^- D^{*0} (D^{*0}\to \pi^0 D^0, D^0\to K^-\pi^+)$ where the selection of charged particles is identical, and the total uncertainty on the efficiencies for the two $\pi^0$s is estimated to be 7.7\%.
We assess a systematic uncertainty associated with the $\pi^+$ identification efficiency using the $D^{*+}\to D^0 (\to K^-\pi^+) \pi^+$ decay.
The data-simulation PID efficiency ratio is $R=0.962\pm0.001$ per $\pi^+$.
A correction factor $R=1.051\pm0.022$ for the continuum suppression efficiency due to data-simulation discrepancies is determined from control samples of $B^0\to D^{*-} (\to \bar{D}^0 (\to K^+\pi^-) \pi^-) \pi^+$ decays.
The total number of $B\bar{B}$ pairs produced, $(198\pm3)\times10^6$, is obtained from a fit to an event shape variable, the second Fox-Wolfram moment $R_2$, where uncertainties associated with off-resonance subtraction, beam energy shift and spread, and selection efficiency are included.
A systematic uncertainty on the single-candidate selection is evaluated by re-doing the analysis by randomly selecting one $B^0$ candidate per event.
The difference between the two analysis results is taken as a systematic uncertainty.

In order to estimate systematic uncertainties due to signal or background mismodelings, we fit the data by changing the PDF parameters including the yields of $B^0\to a^+_1\pi^-$, $B^0\to K^{*+} \rho^-$, and $B^0\to K^{*+}_0 \rho^-$ within their 1$\sigma$ uncertainties, and then, the uncertainties of the self cross-feed fractions are set to 15\%.
In the systematic uncertainty, we include statistical biases observed in fits to ensembles of simplified simulated experiments.
Slight data-simulation discrepancies are observed in the $\cos \theta_{\rho^\pm}$ distributions for candidates populating the signal $M_{bc}$ sideband that cannot be conclusively attributed to shape or acceptance mismodelings or poorly simulated sample composition.
A systematic uncertainty based on the deviation of results in fits to simulated ensembles that mirror the observed discrepancies covers a broad range of effects in $\cos \theta_{\rho^\pm}$.

\begin{table}[tb]
  \caption{Summary of the systematic uncertainties.}
  \label{tab:syst-error}
  \centering
  \begin{tabular}{lcc}
    \hline \hline
    source & $\mathcal{B}$ [\%] & $f_L$ [\%]\\
    \hline
    Tracking                              & 0.6 & - \\
    Photon and $\pi^0$ selection          & 7.7 & - \\
    PID                                   & 0.8 & - \\
    Continuum suppression                 & 2.1 & - \\
    $N_{B\bar{B}}$                        & 1.5 & - \\
    Single candidate selection            & 2.2 & 0.9 \\
    Signal model                          & 2.4 & 2.0 \\
    Self cross-feed model                 & $^{+2.7}_{-0.9}$ & $<0.1$ \\
    Continuum model                       & 1.3 & 0.7 \\
    $B\bar{B}$ model                      & 2.0 & 2.2 \\
    Peaking background model              & 0.4 & 0.7 \\
    $\cos\theta_{\rho^{\pm}}$ mismodeling & 4.4 & 0.3 \\
    Fit bias                              & 0.9 & 1.0 \\
    Simulation sample size                & 1.0 & 0.2 \\
    \hline 
    Total & ${}^{+10.6}_{-10.3}$ &  $\pm3.4$ \\
    \hline \hline
  \end{tabular}
\end{table} 

\section{Results and summary}
\begin{figure}[t]
  \centering
  \includegraphics[width=1\textwidth]{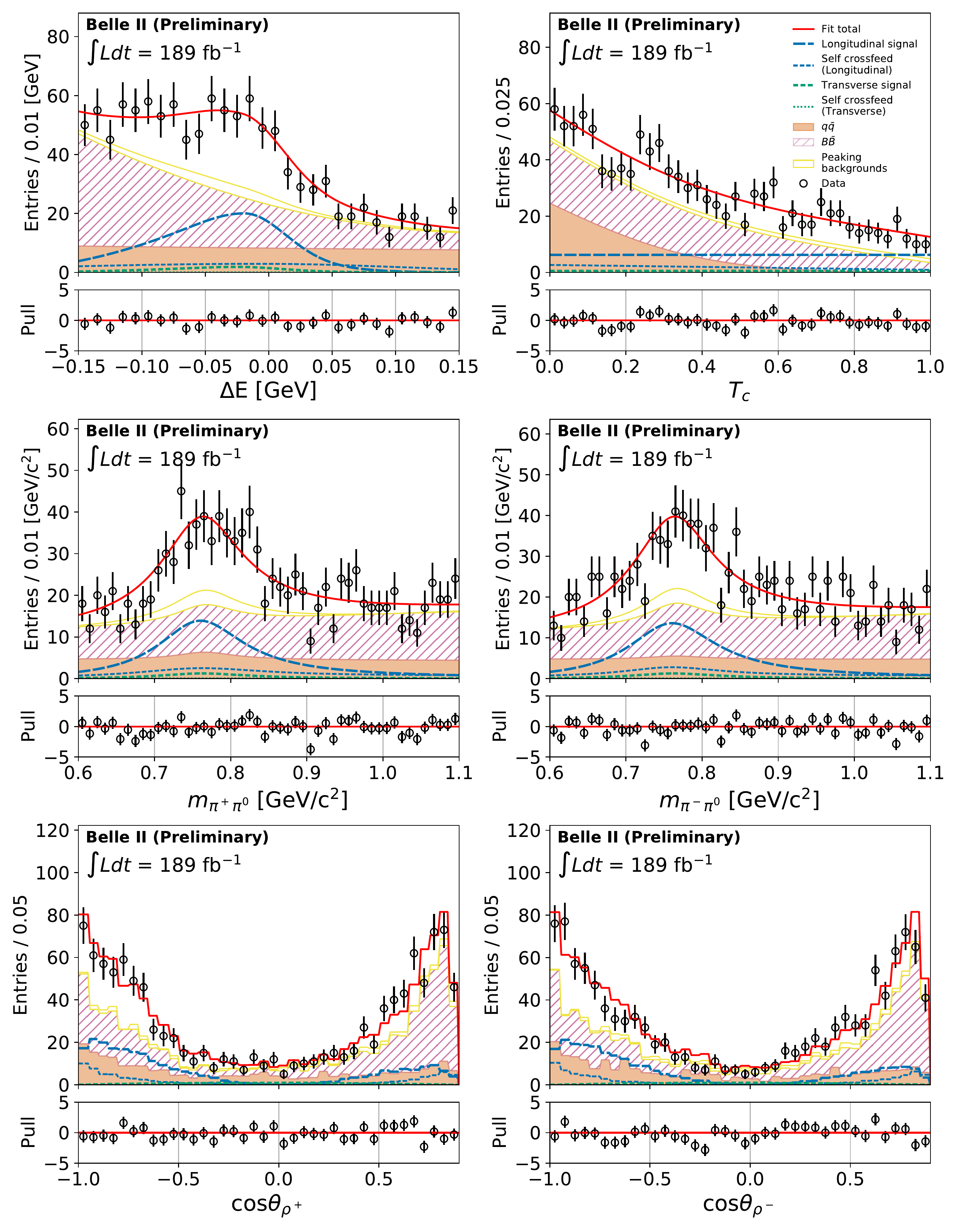}
  \caption{Distribution of (top left) $\Delta E$, (top right) $T_C$, (middle left) $m_{\pi^+\pi^0}$, (middle right) $m_{\pi^-\pi^0}$, (bottom left) $\cos \theta_{\rho^+}$, and (bottom right) $\cos \theta_{\rho^-}$ for the $B^0\to\rho^+\rho^-$ candidates reconstructed in 2019-2012 Belle II data passing through the optimized selection. Fit projections are overlaid.}
  \label{fig:fit}
\end{figure}

We report measurements of the branching fraction and the longitudinal polarization in $B^0\to\rho^+\rho^-$ decays based on a 189~fb$^{-1}$ data sample containing $(198 \pm 3) \times 10^6 B\bar{B}$ pairs.
The distributions in data with fit results overlaid are shown in Figure~\ref{fig:fit}.
The resulting branching fraction and longitudinal polarization are
\begin{eqnarray*}
  \mathcal{B}(B^0\to\rho^+\rho^-) &=& [2.67\pm0.28\,(\mathrm{stat})\,\pm0.28\,(\mathrm{syst})] \times 10^{-5},\\
  f_L &=& 0.956\pm0.035\,(\mathrm{stat})\,\pm 0.033\,(\mathrm{syst}),
\end{eqnarray*}
with a correlation coefficient of $-0.28$. 
The signal yields determined by the fit are summarized in Table~\ref{tab:fit-result}.
We confirm longitudinal polarization dominance in $B^0 \to \rho^+\rho^-$ decays, which allows precise time-dependent {\it CP} violation measurements in this decay mode.
This result is consistent with previous results~\cite{Belle:2015xfb, BaBar:2007cku, ParticleDataGroup:2022zzz}, and will be used to constrain penguin pollution together with $B^+ \to \rho^+ \rho^0$~\cite{Belle:2003lsm, BaBar:2009rmk, Belle-II:2022ksf} and $B^0 \to \rho^0 \rho^0$~\cite{BaBar:2008xku, Belle:2012ayg, LHCb:2015zxm} decays and to extract the CKM angle $\phi_2$.

\begin{table}[ht]
  \caption{Fit results for signal yields and efficiencies. The uncertainties are statistical.}
  \label{tab:fit-result}
  \centering
  \begin{tabular}{lcc}
    \hline \hline
    Polarization & $N$ & $\epsilon$ [\%] \\
    \hline
    Longitudinal  & $234.6 {}^{+23.7}_{-22.8}$ &  $4.76$ \\
    Transverse    & $21.3 {}^{+18.7}_{-17.3}$  &  $9.53$ \\
    \hline \hline
  \end{tabular}
\end{table}
